\documentclass[12pt]{iopart} 
\usepackage{float}
\usepackage{iopams}  
\usepackage{graphicx}
\begin{document}
\title{Nonlinear dynamics of non-neutral Maxwellian plasma in external trapping field}
\author{A.R.~Karimov$^{1,2^*}$, A.A. Dementev$^{2}$, S.A. Terekhov$^{2}$ }
\address{$^1$Institute for High Temperatures,
Russian Academy of Sciences, Izhorskaya 13/19, Moscow 127412, Russia\\
$^2$National Research Nuclear University MEPhI, Kashirskoye shosse 31, Moscow, 115409, Russia\\
$^*$email: arkarimov@mephi.ru}
\begin{abstract}
Nonlinear dynamics of collisionless non-neutral plasma in an external electrical trapping field is considered. 
Time-dependent solution of the nonlinear Vlasov-Poisson equations are obtained. 
The influence of initial conditions on the dynamics of charged layer is discussed.
\end{abstract}


\section{Introduction}

The evolution of collisionless, non-neutral plasma in an electrostatic or magnetic trap is an interesting case of complex physical system far from equilibrium where depending on the initial conditions and parameters of the external trapping fields some dynamical structures are formed in phase space by the nonlinear wave-particle interactions \cite{Dv_1974} -- \cite{Kr_1900}. It means that there are many ways for evolution of the non-neutral plasma system from the initial state far from equilibrium to a state, where the distribution function practically ceases to change with time.

One of the possible way is a dynamics via the formation of local equilibrium state formed in the self-generated and external holding fields (see, for example, \cite{Kr_1900, anderegg}). In the present paper, we are going to study this case in two-dimensional geometry for the cloud of the single charged particles trapping by symmetric quadratic electrostatic field. In order to investigate the evolution of such system, we shall construct exact nonstationary solution of the two-dimensional Vlasov-Poisson equations, which has a local equilibrium form. On the other hand, such solution belongs to the class of exact solutions of the fully nonlinear equations describing a non-neutral, warm plasma. Using this solution, we will discuss the physical conditions for the realization of local equilibrium states in charged media of the type under consideration.


\section{Formulation of the kinetic problem}

We consider the dynamics of two-dimensional slab of non-neutral, non-relativistic, collisionless medium constituted by particles with the same mass $m$ and charge $q$, which is confined in the quadratic potential
\begin{equation}
\Phi_{ext}(x,y)=k_xx^2+k_y y^2\/,
\label{p_2}
\end{equation}
where $k_x$ and $k_y$ are the constant components of the external focusing parameter in the $Oxy$ plane. Such dependence is typical for Paul, Kingdon and Penning traps. The dimensionless Vlasov-Poisson equations for the one-particle distribution function  
$f=f(t, x, y, v_x, v_y)$, which describe the electron plasma in the electrostatic approximation, are
\begin{equation} 
\frac{\partial f}{\partial t}+v_x \frac{\partial f }{\partial x}+v_y \frac{\partial f }{\partial y}
+\left(E_x-\frac{\partial {\Phi}_{ext}}{\partial x}\right)\frac{\partial f }{\partial v_x}
+\left(E_y-\frac{\partial {\Phi}_{ext}}{\partial y}\right)\frac{\partial f }{\partial v_y}=0\/,
\label{1_nature}
\end{equation}
\begin{equation}
\nabla\cdot \mathbf{E}=4\pi n,
\label{2_nature}
\end{equation}
where $n$ is the density of particles forming a charged medium, $\mathbf{E}=-\nabla\Phi$ and $\Phi(t,x,y)$ are the self-consistent electric field and potential, respectively.
The density, time and space have been normalized by the initial electron density $n_0$, the inverse plasma frequency 
${\omega}_{b} = {(4\pi n_0 e^2/{m}_{e})}^{1/2}$ and the inverse Debye length 
${\lambda}_{b}={({T}_{e}/4\pi n_0 e^2)}^{1/2}$, where ${m}_{e}$ the mass, $e$ the charge, ${T}_{e}$ the temperature 
of electrons.
The velocity and self-consistent potential are
normalized by ${\omega}_{b}{\lambda}_{b}$ and ${T}_{e}/e$, respectively.

Assume that the system is in the state of local equilibrium. Then we look for the partial solution of time-dependent system of Vlasov-Poisson equations (\ref{1_nature})--(\ref{2_nature}) in the local-equilibrium form
\begin{equation} 
f(t,x,y,v_x,v_y)=\frac{n}{{\pi T }}\exp\left(-\frac{(\mathbf{v}-\mathbf{V})^2}{T}\right)\/,
\label{4_nature}
\end{equation}
where $(\mathbf{v}-\mathbf{V})^2=(v_x-V_x)^2+(v_y-V_y)^2$ and $n=n(t)$, $T=T(t,x,y)$, $V_x=V_x(t,x,y)$, $V_y=V_y(t,x,y)$ are unknown functions to be determined.

In view of the symmetry of the slab, we set the additional conditions
\begin{equation} 
\left.\frac{\partial\Phi }{\partial x}\right|_{x=0}=\left.\frac{\partial\Phi }{\partial y}\right|_{y=0}=0, \/
\label{6_nature}
\end{equation}
\begin{equation} 
{k}_{x}={k}_{y}=k.
\label{6x1_nature}
\end{equation}
The relations (\ref{2_nature}) and (\ref{6_nature}) are satisfied by the function
\begin{equation}
\Phi (t,x,y)=\frac{n(t)}{2}\left(x^2+y^2 \right),
\label{7_nature}
\end{equation}
which corresponds to the self-generated field
\begin{equation}
\mathbf{E}(t,x,y)=-\nabla\Phi =-n(t)x\mathbf{e}_x-n(t)y\mathbf{e}_y.
\label{8_nature}
\end{equation}


\section{Evolution equations}

Comparing (\ref{p_2}) with (\ref{8_nature}), taking into account (\ref{6x1_nature}), we conclude that for $k>0$ external and self-consistent fields are oppositely directed: this makes it possible to confine the non-neutral plasma. For this condition, 
the substitution of (\ref{4_nature}) into (\ref{1_nature}) leads to
\begin{eqnarray}
{v}^{(0)} \left[\frac{\dot{n}}{n}-\frac{\dot{T}}{T}+\frac{\dot{T}}{T^2}V^2-\frac{2}{T}\left(V_x\dot{V}_x +V_y\dot{V}_y\right)+\frac{2}{T}\left(n-k\right)\left(xV_x+yV_y\right) \right]+\nonumber\\
{v_x}^{(1)} \left[-2\frac{\dot{T}}{T^2}V_x+\frac{2}{T}\dot{V_x}-\frac{2V_x}{T}\frac{\partial V_x}{\partial x}-\frac{2V_y}{T}\frac{\partial V_y}{\partial x}-\frac{2(n-k)x}{T} \right]+\nonumber\\
{v_y}^{(1)} \left[-2\frac{\dot{T}}{T^2}V_y+\frac{2}{T}\dot{V_y}-\frac{2V_x}{T}\frac{\partial V_x}{\partial y}-\frac{2V_y}{T}\frac{\partial V_y}{\partial y}-\frac{2(n-k)y}{T} \right]+\nonumber\\
v_x v_y \left[ \frac{2}{T}\frac{\partial V_y}{x}+\frac{2}{T}\frac{\partial V_x}{y} \right]+
{v_x}^{(2)}\left[\frac{\dot{T}}{T^2}+\frac{2}{T}\frac{\partial V_x}{x} \right]+\nonumber\\
{v_y}^{(2)}\left[\frac{\dot{T}}{T^2}+\frac{2}{T}\frac{\partial V_y}{y} \right]=0.
\label{77_nature}
\end{eqnarray}
This relation is valid for any $v_x$ and $v_y$. Thus, the coefficients of each power of $v_x^s v_y^l$ should be equal to zero for any $l$ and $s$.  Nullifying the coefficients of $v_x v_y$, ${v_x}^{(2)}$ and ${v_y}^{(2)}$ in equation (\ref{77_nature}), we obtain
\begin{equation}
\frac{\partial V_y}{\partial x}=-\frac{\partial V_x}{\partial y},
\label{aa4_nature}
\end{equation}
\begin{equation}
\frac{\partial V_x}{\partial x}=-\frac{1}{2}\frac{\dot{T}}{T}, \hspace{9mm}
\frac{\partial V_y}{\partial y}=-\frac{1}{2}\frac{\dot{T}}{T}.
\label{aa5_nature}
\end{equation}
In the simplest case, if $T$ does not depend on the coordinate, from the relations (\ref{aa5_nature}) it follows that
\begin{equation}
V_x(t,x)=-\frac{1}{2}\frac{\dot{T}}{T}x+g(y,t)\/,
\label{10x_nature}
\end{equation}
\begin{equation}
V_y(t,y)=-\frac{1}{2}\frac{\dot{T}}{T}y+h(x,t)\/.
\label{11x_nature}
\end{equation}
The equations (\ref{10x_nature}), (\ref{11x_nature}) must satisfy the condition (\ref{aa4_nature}). Now we show that this relation will be satisfied when
\begin{equation} 
h(x,t)=Cx+D_1(t)\/,
\label{zz_nature}
\end{equation}
\begin{equation} 
g(y,t)=-Cy+D_2(t)\/,
\label{z_nature}
\end{equation}
where $C$ is arbitrary constant, $D_1(t)$, $D_2(t)$ are some functions. Indeed, since
\begin{equation}
\frac{\partial V_x}{\partial y}=\frac{\partial g}{\partial y}, \hspace{4mm}
\frac{\partial V_y}{\partial x}=\frac{\partial h}{\partial x} \hspace{4mm}
,\hspace{4mm}
\frac{\partial g}{\partial y} = -\frac{\partial h}{\partial x},
\label{z1z_nature}
\end{equation}
then relation (\ref{aa4_nature}) is satisfied.

Substituting the relations (\ref{10x_nature}) and (\ref{11x_nature}) into equation (\ref{77_nature}), we can equate to zero 
the coefficients before $x^s y^l$ for any $l$ and $s$. Thus, we ensure that this equation is satisfied for arbitrary $x$ and $y$. 
As a result, we obtain the relations
\begin{equation}
\frac{d}{dt}\left(\frac{\dot{T}}{T} \right)-\frac{1}{2}\left(\frac{\dot{T}}{T} \right)^2 +
2n-2k=0,
\label{nontriv23}
\end{equation}
\begin{equation}
\frac{\dot{n}}{n}-\frac{\dot{T}}{T}+\frac{\dot{T}}{T^2}(D_1^2+D_2^2)+\frac{2}{T}D_1\dot{D_1}+\frac{2}{T}D_2\dot{D_2}=0,
\label{nontriv4}
\end{equation}
\begin{equation}
\dot{D_1}=\frac{1}{2}\frac{\dot{T}}{T}D_1,
\label{nontriv19}
\end{equation}
\begin{equation}
\dot{D_2}=\frac{1}{2}\frac{\dot{T}}{T}D_2.
\label{nontriv20}
\end{equation}
Equations (\ref {nontriv19}), (\ref {nontriv20}) have the following solutions
\begin{equation}
D_1=D_a{T}^{\frac{1}{2}},
\label{nontriv22}
\end{equation}
\begin{equation}
D_2=D_b{T}^{\frac{1}{2}},
\label{nontriv21}
\end{equation}
where $D_a$, $D_b$ are constants. For simplicity we put $D_a=D_b=D_0$. In this case we have
\begin{equation}
D_1=D_2=D_0{T}^{\frac{1}{2}}.
\label{nontriv22b}
\end{equation}
By substituting (\ref{nontriv22b}) into (\ref{10x_nature}) and (\ref{11x_nature}), for initial moment of time, we obtain
\begin{equation}
D_0 = V_x (t=0) {T(t=0)}^{-\frac{1}{2}} = V_y(t=0) {T(t=0)}^{-\frac{1}{2}}.
\label{144_nature}
\end{equation}
Thus, $D_0$ determines the initial distribution of hydrodynamic velocities.

Taking into account (\ref {nontriv22}), from the relation (\ref {nontriv4}) we get
\begin{equation}
\frac{\dot{n}}{n}+\frac{\dot{T}}{T}\left(4{D_0}^{2}-1 \right)=0.
\label{nontriv24}
\end{equation}
The solution of this equation is
\begin{equation}
n=n_0{T}^{1-4{D_0}^{2}}.
\label{nontriv25}
\end{equation}
For simplicity, we put the constant of integration equal to $n_0$. This solution allows us to exclude the density $n(t)$ from the equation (\ref {nontriv23}), which as a result becomes
\begin{equation}
\frac{d^2 u}{dt^2}-\frac{1}{2}\left[\frac{du}{dt} \right]^2 +
2n_0 {e}^{u\alpha}-2k=0\/,
\label{nontriv26}
\end{equation}
here we use the notation
\begin{equation}
u=\ln T, \hspace{9mm} \alpha=1-4{D_0}^{2}\/.
\label{nontriv27}
\end{equation}
By using (\ref{10x_nature}), (\ref{11x_nature}), (\ref{nontriv25}) and (\ref{nontriv27}), we express the parameters in the distribution (\ref{4_nature}) in explicit form through the function $u$ and parameter $\alpha$:
\begin{equation}
\eqalign{n=n_0e^{u\alpha}, \hspace{9mm} T= e^u, \cr
V_x=-\frac{1}{2}\frac{du}{dt}x+\sqrt{\frac{1-\alpha}{4}} {e}^{\frac{u}{2}},\hspace{9mm}
V_y=-\frac{1}{2}\frac{du}{dt}y+\sqrt{\frac{1-\alpha}{4}} {e}^{\frac{u}{2}}\/.}
\label{nontriv28}
\end{equation}
Thus, proceeding from these relations and (\ref{nontriv26}) we can determine the evolution of the system worked out.


\section{Dynamics of a charged slab}

The physical properties of a non-neutral plasma are determined by the dynamic properties of the solutions of the equation 
(\ref {nontriv26}), depending on the initial conditions. 
In the absence of heating or cooling sources in the system we have to put
\[
\left.\frac{\partial u }{\partial t}\right|_{t=0}=0\/,
\]
which corresponds to immobile non-neutral medium in initial moment of time, and to
consider changing of relevant parameters in the local equilibrium distribution (\ref{4_nature})
for different initial temperatures $u (t = 0) = u_0$, the values of the external focusing parameter $k$ 
and the parameter $\alpha$.

As is seen from the relation (\ref{nontriv28}), the parameter $\alpha$, associated with the functions $n$ and $V_x$, $V_y$, defines the initial distribution of the hydrodynamic velocities.
In the case of $\alpha=1$, the hydrodynamic velocities at the initial moment of time are equal to zero. 
Therefore, it is convenient to use this value in further calculations.

\begin{figure}[H]
\begin{minipage}[h]{0.47\linewidth}
\begin{center}
\includegraphics[width=7.8cm]{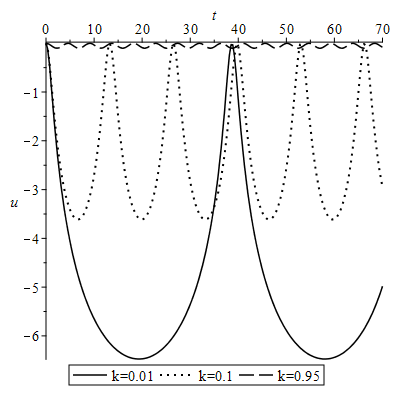}  \\(a)
\end{center}
\end{minipage}
\hfill
\begin{minipage}[h]{0.47\linewidth}
\begin{center}
\includegraphics[width=7.8cm]{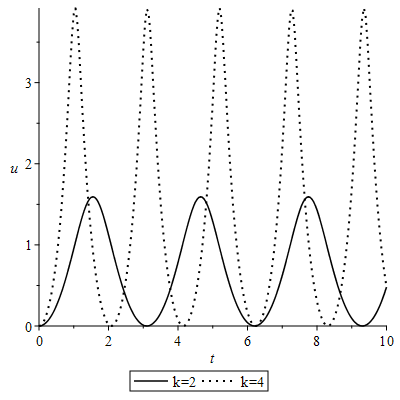}  \\(b)
\end{center}
\end{minipage}
\caption{Evolution of the temperature dependence $u(t)$ for $\alpha=1$, $ u'(t=0)=0 $, $u_0=0$ under different values of the parameter $k$.
\label{f2}}
\end{figure}

Fig. \ref{f2} shows the evolution of the function $u(t)$ associated with density, temperature and velocity by the relations (\ref{nontriv28}) for different values of the focusing parameter $0<k<4$ and the initial conditions $u'(t=0)=u(t=0)=0$, which corresponds to a immobile non-neutral plasma with nonzero initial temperature $T(t=0)=1$.

In the case of small values $k\leq 1$ (Fig. \ref{f2} (a)), when $u<0$, the charged layer is expanded accompanied by nonlinear oscillations of the function $u(t)$. As is seen from the above dependences, increase in the parameter $k$ leads to increase in the oscillation frequency. 
However, according to (\ref{nontriv26}) and (\ref{nontriv28}), these oscillations are most strongly manifested in the velocity field, while the other parameters show insignificantly sensitivity to oscillations of $u$. 
The interval $1 \leq k \leq 4$ corresponds to confinement of the plasma. For $k=1$ the function $u$, and hence the temperature $T$ and the density $n$, do not change with time, the velocities $V_x$ and $V_y$ become equal to zero. 
A further increase in $k$ leads to the emergence of strong nonlinear oscillations in all parameters. 
For $k>4$ the distribution $u(t)$ exhibits a singular behavior, which is due to the dynamics of the velocity field.

Thus, from the point of view of plasma confinement, the value $k = 1$ is of the greatest interest, since here the parameters in the distribution (\ref{4_nature}) remain constant.

\begin{figure}[H]
\begin{minipage}[h]{0.47\linewidth}
\begin{center}
\includegraphics[width=7.8cm]{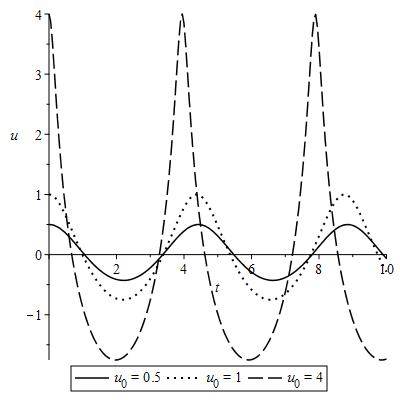}  \\(a)
\end{center}
\end{minipage}
\hfill
\begin{minipage}[h]{0.47\linewidth}
\begin{center}
\includegraphics[width=7.8cm]{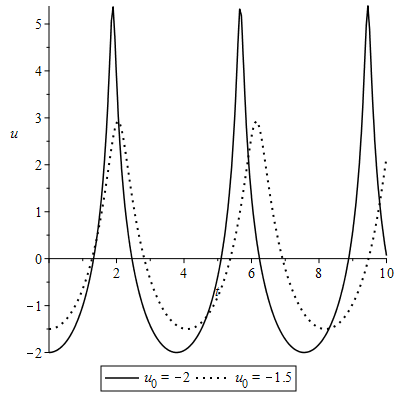}  \\(b)
\end{center}
\end{minipage}
\caption{Evolution of the temperature dependence $u(t)$ for $\alpha=1$, $ u'(t=0)=0 $, $u_0=0$ under different values of the initial temperature $u_0$.
\label{f4}}
\end{figure}

For this case Fig. \ref {f4} shows the influence of the initial temperature for $u'(t=0)=0$ on the dynamics of the resulting oscillations in the interval 
\[
- 2 \leq u_0 \leq 4.
\]

\begin{figure}[H]
\begin{minipage}[h]{0.47\linewidth}
\begin{center}
\includegraphics[width=7.8cm]{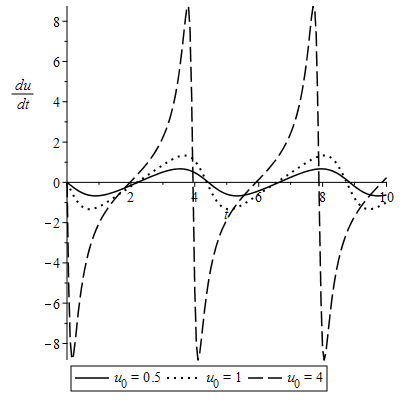}  \\(a)
\end{center}
\end{minipage}
\hfill
\begin{minipage}[h]{0.47\linewidth}
\begin{center}
\includegraphics[width=7.8cm]{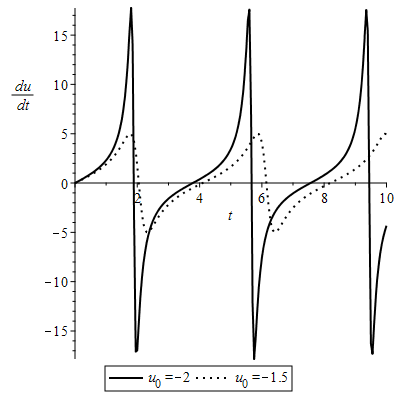}  \\(b)
\end{center}
\end{minipage}
\caption{Dynamics of the first derivative of the function $u(t)$ for $\alpha=1$, $k=1$, $u'(t=0)=0$ under different values of the initial temperature $u_0$.
\label{f6}}
\end{figure}

The values of the initial temperature greater than $u_0 = 4$ lead to the emergence of a singularity. This is due to rollover of the velocity profile, that is shown in Fig. \ref {f6}. These graphs show that for $u_0 \geq 1$ we can observe plasma confinement, accompanied by the emergence of persistent nonlinear oscillations.

\begin{figure}[H]
\begin{minipage}[h]{0.45\linewidth}
\begin{center}
\includegraphics[width=6.cm]{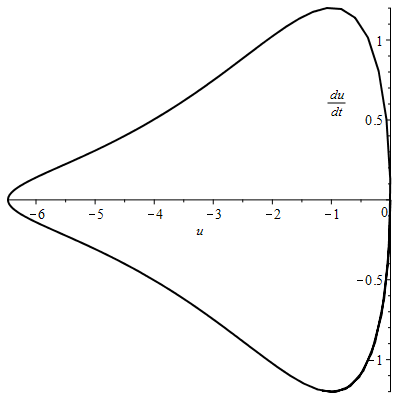}  \\(a)
\end{center}
\end{minipage}
\hfill
\begin{minipage}[h]{0.45\linewidth}
\begin{center}
\includegraphics[width=6.cm]{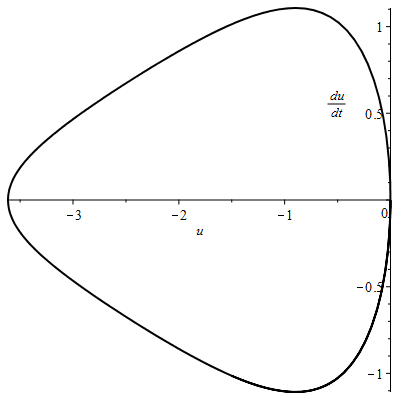}  \\(b)
\end{center}
\end{minipage}
\vfill
\begin{minipage}[h]{0.45\linewidth}
\begin{center}
\includegraphics[width=6.cm]{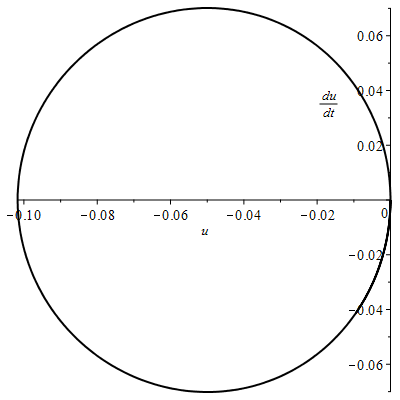}  \\(c)
\end{center}
\end{minipage}
\hfill
\begin{minipage}[h]{0.45\linewidth}
\begin{center}
\includegraphics[width=6.cm]{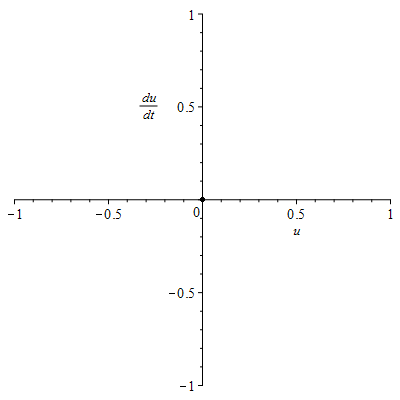}  \\(d)
\end{center}
\end{minipage}
\vfill
\begin{minipage}[h]{0.45\linewidth}
\begin{center}
\includegraphics[width=6.cm]{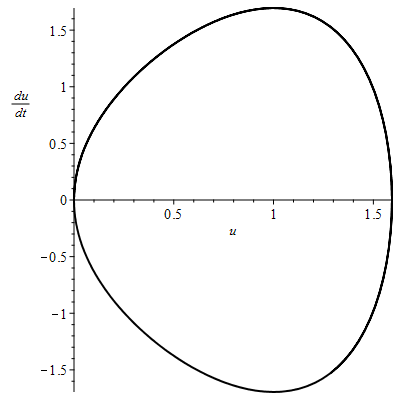}  \\(e)
\end{center}
\end{minipage}
\hfill
\begin{minipage}[h]{0.45\linewidth}
\begin{center}
\includegraphics[width=6.cm]{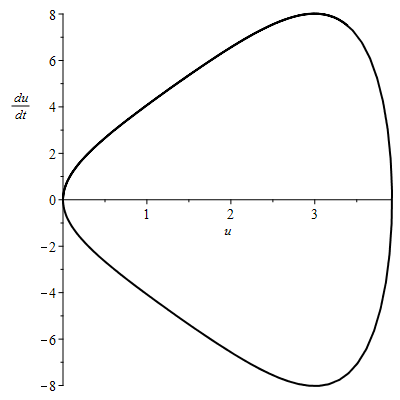}  \\(f)
\end{center}
\end{minipage}
\caption{Phase portraits for $\alpha=1$, $ u'(t=0)=0 $, $u_0=0$ under different values of the parameter $k$:
(a) $k=0.01$; (b) $k=0.1$; (c) $k=0.95$; (d) $k=1$; (e) $k=2$; (f) $k=4$. 
\label{f103}}
\end{figure}

Thus, the dependencies presented in Figs. \ref {f2} - \ref {f6}, point out to the existence of a range of $k$ and $u_0$, in which plasma is confined. For example, we specify a neighborhood near the point $k = 1, u_0 = 0$. Apparently, such states can be identified with stable, long-lived states that can be obtained experimentally if the distribution of the form (\ref {4_nature}) has already been established in the system. The change in the nature of the undamped nonlinear oscillations depending from $k$ is illustrated by the phase portraits shown 
in Fig. \ref {f103}.

As is seen, an increase in the value $k$ from the interval $0<k \leq1$ leads to the appearance of bounded solutions and a decrease in the phase volume vanishing in the limit for $k=1$ (Fig. \ref {f103} (d)) . The values $k>1$ lead to a gradual increase in the phase volume and the emergence of singular behavior.

Now we consider the dynamics of the nonlinear oscillations in the temperature dependence $u(t)$ under the variation of the parameter $\alpha$.

\begin{figure}[H]
\begin{center}
\includegraphics[width=16.5cm]{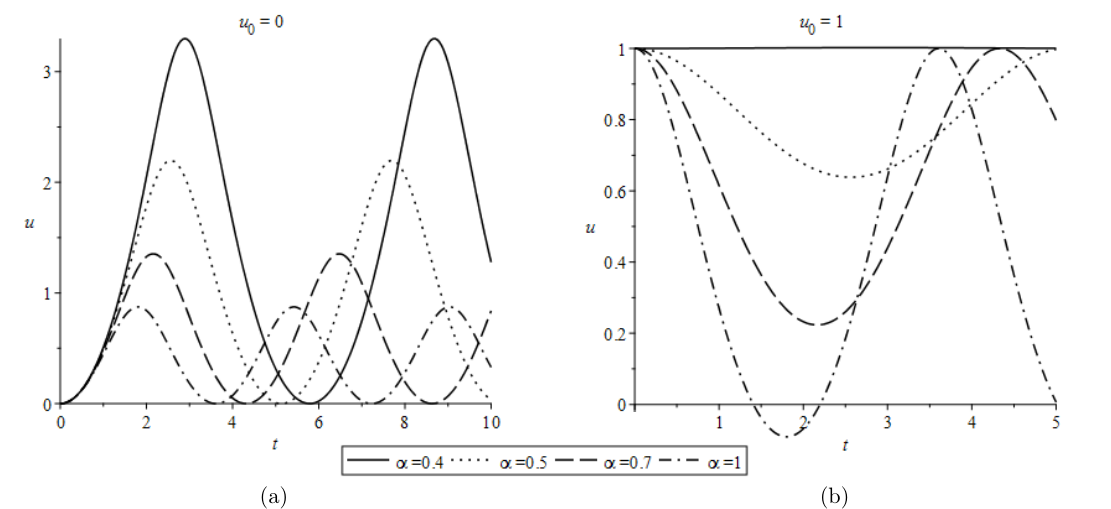}  
\end{center}   
\caption{Influence of the parameter $\alpha$ on the dynamics of the oscillations of the function $u(t)$ for $k=1.5$, $u'(t = 0)=0$ under different values of the initial temperature $u_0$.
\label{a_2}}
\end{figure}

Fig. \ref{a_2} (a)  shows the plots of the function $u(t)$ for various values of the parameter $0.4 \leq \alpha \leq 1$ and fixed initial conditions $k=1.5$, $u'(t=0)=u(t=0)=0$. 
As is seen from these dependencies, the amplitude of the oscillations of the function $u(t)$ increases and shows a tendency to a singular behavior when the parameter $\alpha \rightarrow 0$. 
This leads to the fact that the range of values of the parameter $k$, where the charge layer can be confined, becomes smaller.
For example, if $\alpha=0.4$, this range decreases to $1 \leq k \leq 1.5$. 
In the case of $\alpha>1$, the values of the components of hydrodynamic velocities have an imaginary part.

By setting the value of initial temperature $u_0=1$ for $0.4 \leq \alpha \leq 1$, $k=1.5$, we obtained the set of graphs 
presented in Fig. \ref{a_2} (b).
The comparison of this dependences with dependences in Fig. \ref{a_2} (a), makes it possible to investigate the influence of the initial temperature on the dynamics of the oscillations of the function $u(t)$ for various $\alpha$: we see that the character 
of the amplitude variation for different $u_0$ and $\alpha$ can differ.
Since for the graphs shown in Fig. \ref{a_2} (b), the decrease in the parameter $\alpha$ leads to a decrease in the amplitude of the oscillations of the function $u(t)$, then it becomes possible to select the conditions under which the function $u(t)$ does not change with time. 
In our case, this is achieved for $u_0=1$, $\alpha=0.4$, $k=1.5$. 
Thus, by varying the parameters $\alpha$ and $k$ with initial temperature $u_0$, for which a decrease in parameter 
$\alpha$ leads to a decrease in the amplitude of the oscillations of the function $u(t)$, we can find a conditions for the constancy
of temperature when the charged layer is confined.

\section{Conclusion}

In this paper we have considered the dynamics of non-neutral single-component plasma in the external electrostatic field with a quadratic potential for a two-dimensional geometry. As a particular solution of the Vlasov-Poisson equations, we used the distribution of the local-equilibrium form (\ref{4_nature}), whose macroscopic parameters are described by the dependencies (\ref{nontriv27}) and (\ref{nontriv28}) through the function u(t), which is the logarithm of temperature, and the parameter $\alpha$.
The function u(t) satisfies the equation (\ref{nontriv26}), which describes a series of non-stationary bounded solutions, depending on the initial temperatures, the external focusing parameter $k$ and the parameter $\alpha$. Analysis of this equation point out to existence of a range of initial conditions, in which nonstationary confinement of a non-neutral plasma is possible. In fact, some of the presented results on the evolution of initially confined non-neutral plasmas are similar to phenomena predicted by the latter in Refs. \cite{Kr_1900} -- \cite{KYS_2016}. 

The two-dimensional model considered here describes a particular case of the evolution of a charged medium, since real systems have a higher spatial dimension, even in the absence of plasma confinement and dissipation processes. On the other hand, our exact solutions, that describe possible strongly nonlinear final states, can give a basic idea of the formation of time-dependent structures in a real system.


\end{document}